\documentclass[12pt]{article}

\usepackage{array}
\usepackage{amssymb,amsmath}
\usepackage{fullpage}
\usepackage[dvips]{graphicx}

\usepackage{amssymb,amsmath,eepic,epic,colordvi,mathrsfs}
\usepackage{float}
\usepackage{fullpage}

\newcommand {\ignore} [1] {}

\def\newtheorems{\newtheorem{theorem}{Theorem}[section]

\newtheorem{prop}[theorem]{Proposition}
\newtheorem{lemma}[theorem]{Lemma}

\newtheorem{definition}[theorem]{Definition}}

\def\squarebox#1{\hbox to #1{\hfill\vbox to #1{\vfill}}}
\newcommand{\qed}{\hspace*{\fill}
\vbox{\hrule\hbox{\vrule\squarebox{.667em}\vrule}\hrule}\smallskip}

\newtheorems

\newcommand{\Tp}{\mbox{$T_P$}}
\newcommand{\Th}{\mbox{$T_H$}}

\newcommand{\Dn}{\mbox{D\_num}}
\newcommand{\Fp}{\mbox{$F_P$}}
\newcommand{\Fh}{\mbox{$F_H$}}

\newcommand{\remnum}{\mbox{\it removal\_num}}
\newcommand{\delnum}{\mbox{\it deliver\_num}}

\newcommand{\prerem}{\mbox{pre\_rem}}

\newcommand{\tp}{\mbox{\it tp}}

\newcommand{\removeit}{{\it remove}}

\newcommand{\fp}{\mbox{\it fp}}

\newcommand{\fh}{\mbox{\it fh}}
\newcommand{\itth}{\mbox{\it th}}

\newcommand{\Rn}{\mbox{R\_num}}

\newcommand{\IF}{{\sf if}}
\newcommand{\REPEAT}{{\sf repeat}}
\newcommand{\UNTIL}{{\sf until}}
\newcommand{\RETURN}{{\sf return}}

\newcommand{\ELSE}{{\sf else}}

\newcommand{\THEN}{{\sf then}}

\newcommand{\TRUE}{\mbox{\sf true}}
\newcommand{\FALSE}{\mbox{\sf false}}

\newcommand{\Boolean}{\mbox{\it Boolean}}

\newcommand{\End}{{\it end}}
\newcommand{\Begin}{{\it begin}}

\newcommand{\Event}{\mbox{\it Event}}

\newcommand{\Number}{\mbox{\it Number}}

\newcommand{\enq}{\mbox{\it enq}}
\newcommand{\deq}{\mbox{\it deq}}

\newcommand{\Val}{\mbox{\it Val}}

\newcommand{\postman}{\mbox{\it postman}}

\newcommand{\homeowner}{\mbox{\it home-owner}}

\newcommand{\positive}{\mbox{\it positive}}
\newcommand{\negative}{\mbox{\it negative}}
\newcommand{\deliverit}{\mbox{\it deliver}}

\newcommand{\checkit}{\mbox{\it check}}

\newcommand{\II}{\mbox{I \kern -4.5pt I}}
\newcommand{\III}{{I \kern -4.5pt I \kern -4.5pt I}}

\newcommand{\R}{\mbox{\sf R}}

\newcommand{\namedref}[2]{\hyperref[#2]{#1~\ref*{#2}}}

\newcommand{\itrn}{\mbox{\it rn}}

\newcommand{\itdn}{\mbox{\it dn}}

\begin{document}

\title{On the mailbox problem}
\author{Uri Abraham and Gal Amram \\ Departments of Mathematics and Computer
Science,\\ Ben-Gurion University, Beer-Sheva, Israel.}

\maketitle

\begin{abstract}
The Mailbox Problem was described and solved by Aguilera, Gafni, and Lamport in \cite{AGL2} with an algorithm that uses two flag registers that carry 14 values each. An interesting problem that they ask is whether there is a mailbox algorithm with smaller flag values. We give a positive answer by describing a mailbox algorithm with 6 and 4 values in the two flag registers.

\end{abstract}

\section{Introduction: the mailbox problem}
\label{PDqa}	

The Mailbox Problem is a theoretical synchronization problem
that arises from analyzing the situation in which a processor must cater to occasional requests from some
device. The problem,
as presented (and solved) in \cite{AGL2} requires the implementation of three operations: \deliverit, \checkit, and \removeit. The device executes a \deliverit\ operation whenever it wants to get the processor's
attention, and the processor executes from time to time \checkit\ operations to find out if there are
any unhandled device requests. After receiving a positive answer for its \checkit\ operation the processor
executes a \removeit\ operation to find-out the nature of the request and to clear the interrupt controller.
It is required that a \checkit\ operation $C$ returns a positive answer if and only if the number of \deliverit\ occurrences that precede $C$ is strictly greater than the number of \removeit\ operations executed
before $C$. The Mailbox Problem is to design a \deliverit/\checkit/\removeit\ algorithm in which
the \checkit\ operation is as efficient as possible, namely that it employs
bounded registers (called ``flags'') that are as small as 
possible. 

\begin{figure}[t]
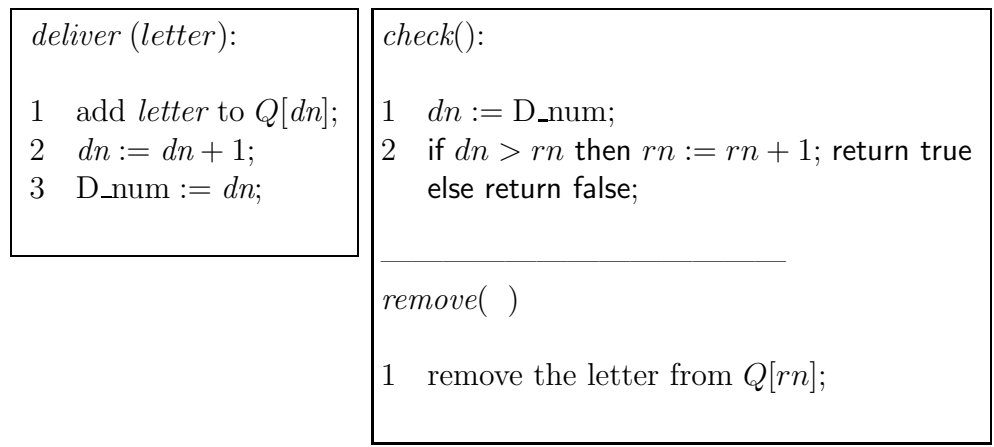

\fbox{
\begin{minipage}[t]{\columnwidth}
\begin{tabbing}
***\=**\=**\=**\=**\=**\=*\=*\=*\=\kill

\deliverit\ $(letter)$:\ \\

\\

1\> add {\it letter} to $Q[\itdn]$;\\

2 \> $\itdn := \itdn + 1$;\\

3  \> $\Dn := \itdn$;\\
 
\end{tabbing}

\end{minipage}
}
\hspace{0mm}
\begin{minipage}[t]{\columnwidth}

\fbox{\begin{minipage}[t]{\columnwidth}

\begin{tabbing}
***\=**\=**\=**\=**\=**\=*\=*\=*\=\kill

\checkit $()$:\ \\
\\

1 \> $dn := \Dn$;\\

2\> \IF\ $dn > rn$\ \THEN\ $rn := rn +1$; \RETURN\ \TRUE\\

\> \ELSE\ \RETURN\ \FALSE;
\end{tabbing}

\begin{tabbing}
***\=**\=**\=**\=**\=**\=*\=*\=*\=\kill
---------------------------------------\\

\removeit(\; ) \\

\\

1 \>  remove the letter from $Q[rn]$;\\

\end{tabbing}

\end{minipage}
}

\end{minipage}

\caption{The unbounded  Mailbox Algorithm. Local variable $dn$ of the postman, and local variable $rn$
of the home-owner are initially 0.}

 \label{fig:ub}

\end{figure}

In \cite{AGL2} the problem is presented first informally
  by means of a story involving two processes, a \postman\ (which is the device) and a 
{\em home owner} (the processor), in which the postman
delivers its letters, and the owner removes them one by one every time she approaches the mailbox.
The problem is to find an algorithm that ensures that the home owner approaches her mailbox if and only if 
it is nonempty. The \checkit\ function  tells the home-owner whether the mailbox
is empty or not, and she approaches her mailbox only after receiving a ``nonempty'' response
from a \checkit\ execution. As noted in \cite{AGL2}, 
depending on the assumptions made on the communication between
the device and processor the mailbox problem can be extremely easy or surprisingly difficult. The following
very easy solution (figure \ref{fig:ub}) shows that if the homeowner process can read an unbounded register then
the mailbox problem becomes trivial. In this unbounded algorithm the postman adds its letter to $Q$ (the queue of requests), and then it writes on its \Dn\ register the number of letters so far added. The home-owner, in executing her
\checkit\ operation, reads register \Dn\ to know how many letters were deposited, and determines the number of messages removed so far by consulting her remove-number local variable $rn$,
and then she concludes that the mailbox is nonempty if the number of letters deposited exceeds the number
of letters removed.

Another easy solution to the mailbox problem can be obtained with stronger communication objects. For example,  a simple algorithm is suggested in \cite{AGL2} in which the postman and home-owner employ
a flag at the mailbox. The postman can atomically (in a single step) deliver mail to the box and raise the flag, 
and the owner atomically removes mail from the box and lowers the flag. The mailbox problem 
becomes highly non-trivial when limitations are
imposed on the communication devices. Specifically, Aguilera et al. require in \cite{AGL2}, for efficiency reasons, that
 the mailbox solutions use only the simplest possible means, and the \checkit\ operation (which is possibly invoked
 at higher frequency) should access only a bounded register. As formulated in \cite{AGL2}, the mailbox problem asks for solutions that satisfy the following requirements\footnote{In an interesting note in his list of publications home-page, Lamport tells that when he first thought about this
problem he believed it has no solution under these requirements.}:

\begin{enumerate}
\item Only registers with read/write actions can be employed.
\item Whereas the \deliverit\ and \removeit\ operations are allowed unbounded registers, the home-owner
can only read bounded value registers in \checkit\ operation executions.
\item Moreover, in her \checkit\ operations the home-owner cannot use persistent local variables, that is
variables that retain their values from one invocation of the operation to the following one.

\item The algorithms for the three operations (\deliverit, \checkit, and \removeit) are bounded wait-free.

\end{enumerate}

A solution is presented in \cite{AGL2} in which each of the two processes uses  unbounded and  bounded registers (the bounded registers are called `flags') and the \checkit\ operation (as required) 
decides on the value to return by reading only the bounded flag registers. The algorithm of \cite{AGL2} needs 14 values
in each of the two flag registers, and a question is posed there if leaner solutions exist. We give a positive 
answer here by describing an algorithm in which the flag registers of the postman and 
the home owner carry  6 and 4 values in each of the flag registers; that is 10 values in total as opposed to 28 values in \cite{AGL2}. 

We shall describe now in more details and greater formality the mailbox problem of \cite{AGL2}.
The mailbox problem assumes two serial processes, a \postman\ process and a \homeowner\ process, and their 
mission is to implement three operations: \deliverit(), \checkit(), and \removeit(). \deliverit() takes a letter as parameter, \checkit\ returns a boolean value, and \removeit() returns a letter.
  It is required that
 the algorithm is bounded wait-free, which means that each operation completes before the process executing it has taken $k$ (atomic) steps, for some fixed constant $k$, irrespectively of what steps the other
 processes take. 

  The \postman\ and the \homeowner\ are serial processes which operate concurrently. The \postman\ executes forever the following routine: he gets a letter $\ell$ and (if the letter is addressed to the home owner) he executes the \deliverit$(\ell)$ operation which adds the letter to the owner's mailbox. So it is quite possible that the total number
  of \deliverit\ operations is finite. The \homeowner\ process executes  forever the following routine:

\begin{equation}
\label{FF}
\begin{array}{l}
\REPEAT\\ \ v:= \checkit()\\ \UNTIL\ v=\TRUE;\\ \ \removeit().
\end{array}
\end{equation}

Thus the $\checkit$ operations are executed ad infinitum, although it is possible that only a finite number of them
are positive (return the value \TRUE). 

  The safety property is expressed in \cite{AGL2} by first stating its sequential specification, and then requiring that a linearization
exists which satisfies this sequential specification. This is the well-known approach to linearizability as defined by
 Herlihy and Wing in \cite{Linearizability}. The following is the formulation in \cite{AGL2} for the sequential specification:
 \begin{quote}

If the owner and postman never execute concurrently, then the value returned by an
execution of \checkit\ is \TRUE\ if and only if there are more deliver than remove executions
before this execution of check.
\end{quote}

To this specification we add the obvious requirement that a queue is implemented, namely that the letters removed are those delivered, and that the letters are removed in the order of delivery. 
The original mailbox paper  \cite{AGL2} mentions
no queues in its algorithms because its authors decided to concentrate on the coordination problem\footnote{In section 2.1 of \cite{AGL2} we read: ``{\it
The remove and deliver procedures are used
only for synchronization; the actual addition and removal of
letters to/from the mailbox are performed by code inserted
in place of the comments. Since it is only the correctness of
the synchronization that concerns us, we largely ignore those
comments and the code they represent}''.}. We prefer however to put the queue in the foreground, since it
seems that the requirement that the home-owner receives the messages of the postman (the device)
and receives them in order is important for the functionality of the system.

We sum-up the requirements of a linear mailbox in Figure \ref{FL}.
\begin{figure}
\fbox{

\begin{minipage}[t]{140mm}
\begin{enumerate}
\item The events are partitioned into \deliverit, \checkit, and \removeit\ events, and are totally ordered by $\prec$ in the order-type of the natural numbers (if not finite).

\item For every \checkit\ event $C$, $\Val(C)\in\{\TRUE,\FALSE\}$. For every \removeit\ event $R$ there is a \checkit\ event $C$ such that $\Val(C)=\TRUE$, $C\prec R$ and  there is no \checkit\ or \removeit\ event $X$ with $C\prec X\prec R$.

\item For every \checkit\ event $C$ let the removal number, $\remnum(C)$, be the number of \removeit\ events $R$ with $R\prec C$, 
and let $\delnum(C)$ be the number of \deliverit\ events $D$ such that $D\prec C$. Then \[\Val(C)=``\remnum(C)<\delnum(C)",\] that is to say the boolean value of $C$ is \TRUE\ iff the number of \deliverit\ events that precede $C$ exceeds the number of letters that were removed by \removeit\ events that precede $C$.

\item If $D_1\prec D_2 \cdots$ and $R_1\prec R_2 \cdots$ are the enumerations in increasing order of the \deliverit\ 
and of the \removeit\ events, then for every $i$ the letter removed by $R_i$ is the letter delivered by $D_i$. 
\end{enumerate}
\end{minipage}
}
\caption{Linear mailbox specification.}
\label{FL}
\end{figure}

As for the liveness requirements, \cite{AGL2} requires that the algorithm is bounded wait-free, which means (see  \cite{6} under the term loop-free, or \cite{3}) that each operation completes before the process executing it has taken $k$ steps, for
some fixed constant $k$.

For communication, the Mailbox Problem as formulated in \cite{AGL2} requires atomic single-writer registers (shared variables). Recall that a register is serial if its read/write events are totally ordered (by the
precedence relation) and the value of any read action is equal to the value of the last write action that
precedes it. A register is atomic if its read/write actions are linearizable into a serial register. That is,
the partially ordered precedence relation has an extension into a total ordering so that the resulting
register is serial. In this paper we assume that all registers are serial. This simplifies somewhat the
presentation of the correctness proof because we do not have to speak about extending the partial order into a linear one,
but it evidently does not limit the applicability of our algorithm which works as well with atomic registers.

For any serial register \R\  we define a function $\omega$ over the read actions of register \R, such that for any read $r$, $\omega(r)$ is the last write action on \R\ that precedes $r$. That is, $\omega(r)< r$ and there is no write action $w$ on \R\
with $\omega(r)< w< r$. Then $r$ and $\omega(r)$ have the same value: $\Val(r)=\Val(\omega(r))$. (To ensure
that $\omega(r)$ is defined on all read actions, we have to assume an initial write event that precedes all read events.)

As we have said, the mailbox algorithm uses both unbounded and bounded registers, but the \checkit\ operation
can access only the bounded registers. Following \cite{AGL2} the bounded registers are called ``flags'',
and so we have the \postman\ flag which we call \Fp\ and the \homeowner\ flag which we call \Fh. The \checkit\ operation only reads these flag registers (and contains no write on any register).

An additional ``access restriction''  is made in \cite{AGL2} for efficiency's sake which requires that the \checkit\ operation uses no persistent
 private variables in a \checkit\ operation. Namely, the owner's decision on whether to approach the mailbox 
or not should depend just on her readings of the $\Fp$ and $\Fh$ values
 and not on any internal information sustained from some previous operation. While one may argue that a small
 persistent variable would not harm the efficiency of the \checkit\ operations, keeping the access
 restriction allows a comparison of the different mailbox algorithms (which obey the same restrictions).
 In fact, if we allow a persistent variables into our \checkit\ algorithm, then the algorithm
 would need just one postman flag register of 6 values and a boolean flag for the home-owner.

\section{The 6/4 mailbox algorithm}
\label{M810}
\begin{figure}[t]
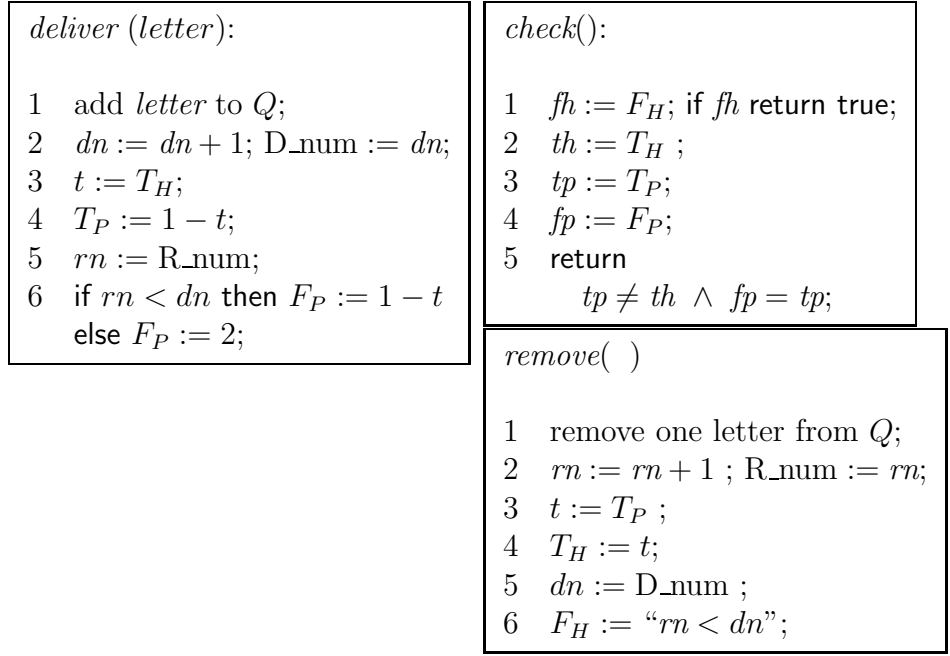


\fbox{
\begin{minipage}[t]{\columnwidth}
\begin{tabbing}
***\=**\=**\=**\=**\=**\=*\=*\=*\=\kill

\deliverit\ $(letter)$:\ \\

\\

1\> add {\it letter} to $Q$;\\

2 \> $\itdn := \itdn + 1$; $\Dn := \itdn$;\\

3  \> $t := \Th$;\\

4 \> $\Tp := 1-t$; \\

5 \> $rn := \Rn$;  \\

6 \> \IF\ $rn<dn$ \THEN\   $\Fp:= 1 - t$\\
\> \ELSE\ $\Fp:= 2$;

\end{tabbing}

\end{minipage}
}
\hspace{0mm}
\begin{minipage}[t]{\columnwidth}
\fbox{
\begin{minipage}[t]{\columnwidth}

\begin{tabbing}
***\=**\=**\=**\=**\=**\=*\=*\=*\=\kill

\checkit $()$:\ \\
\\

1 \> $\fh:= \Fh$; \IF\ \fh\ \RETURN\ \TRUE;\\

2\> $\itth := \Th$ ;\\

3\>  $\tp:=\Tp$;\\

4\>  $\fp := \Fp$;\\

5\> \RETURN\\
\>\>  $\tp\not = \itth\ \wedge\ \fp = \tp$;
\end{tabbing}
\end{minipage}
}

\fbox{
\begin{minipage}[t]{\columnwidth}

\begin{tabbing}
***\=**\=**\=**\=**\=**\=*\=*\=*\=\kill
%----------------------------------\\

\removeit(\; ) \\

\\

1 \>  remove one letter from $Q$;\\

2\> $\itrn := \itrn+1$ ; $\Rn:= \itrn$;\\

3 \> $t:= \Tp$ ;\\

4\> $\Th:= t$;\\

5 \> $dn := \Dn$ ;\\

6 \> $\Fh:= ``\itrn < dn "$;

\end{tabbing}

\end{minipage}
}
\end{minipage}

\caption{The 6/4  Mailbox Algorithm. }

 \label{fig:alg84}

\end{figure}

In this section we define in Figure \ref{fig:alg84} a mailbox algorithm with 6 and 4 values in its two bounded flag registers
 \Fp\ and \Fh. The algorithm uses only serial registers. Registers \Dn, \Tp\ and \Fp\ are written by the postman process, and registers \Rn, \Th\ and \Fh\ are written by the home-owner. Both processes can read these registers, but the \checkit\ procedure only reads the bounded registers: \Fh, \Th,  \Fp\ and \Tp.
 Registers \Dn\ and \Rn\ are unbounded (they carry natural numbers). The bounded registers of the postman process, namely
 \Tp\ and \Fp, are collectively its flag register. Since register \Tp\ carries two values and register \Fp\ three values, the
 combined flag of the postman carries six values. The bounded registers of the homeowner process, \Th\ and \Fh, are
 both boolean, so that there are four values in these two registers which are the flag of the homeowner.

\begin{figure}[t]

\begin{minipage}[t]{\columnwidth} 
 
 \begin{tabular}{ll|l}

 registers & type & initially \\
\hline
{\it of the postman:}\\

\Dn  &   natural number& $0$ \\
     
\Tp   &  $\{ 0,1\}$& $0$ \\

\Fp   &  $\{0,1,2\}$ & $2$ \\
       
\hline\\
{\it of the homeowner:}\\

\Rn &  natural number & $0$\\

\Th & $\{ 0,1\}$ & $0$ \\

\Fh & \Boolean &\FALSE\\
\lasthline
\end{tabular}
 
 \end{minipage}
 \label{table}
 \caption{Registers, their types and initial values.}
 \end{figure}

 We describe the data structures of the different registers in figure \ref{table}.
  \Fp\ values for example are in $\{0,1,2\}$.
The initial values of the registers is also defined in this figure. The initial value of the \Fp\ register for example
is $2$.
 
In addition to the registers, we have the FIFO queue $Q$ which supports two operations: addition of a letter (executed by the postman process), and  removal of a letter (executed by the home owner when $Q$ is nonempty). $Q$ is initially empty. 
 
 The local variables of the algorithm  are as follows. (Variables with unspecified initial values can take
 any initial value.)

\begin{description}
\item[{\it Local variables of \postman}:] $dn$ is a natural number, initially $0$. $rn$ is a natural number, and $t$ is in
$\{0,1\}$.

\item[{\it Local variables of \homeowner}:] Procedure \checkit\ uses variable $fh$ (\Boolean), $th$ and $tp$ (in $\{0,1\}$),
and \fp\ (in $\{0,1,2\}$). The \removeit\ procedure uses $rn$ and $dn$ that are natural numbers, and $t\in \{0,1\}$. Initially $rn=0$.
Local variables of the \postman\ process are obviously different from those of the \homeowner\ even when they have the same name.
\end{description}

In order to ensure that the pseudocode of figure \ref{fig:alg84} is well-understood, we shall go over some
of its instructions, make some simple definitions (that will be used later), and then
we shall explain intuitively some of the main ideas of the algorithm. 

A \deliverit\ operation execution $D$ is an execution of lines 1--6 of that code. It is a high-level event, namely the
set of lower-level actions which are the executions of the code instructions.  Any \deliverit\ execution
is invoked with some letter parameter, and the first line of the code is an enqueue operation
in which this letter is added to $Q$ (the mailbox queue). 

Variable $dn$ (the {\em delivery number}) is initially $0$, so that if $D$ is the $i$'th \deliverit\ operation
execution ($i=1,2,\ldots$) and $dn(D)$ denotes the value of $dn$ after line 2 is executed in $D$, then $dn(D)=i$. Register \Dn\ thus contains the current delivery number.

We shall use this sort of notation $dn(D)$ for other variables as well. We note that in our algorithms any local variable is assigned a value in a unique instruction. So if $v$ is a local variable and $E$ some operation execution that assigns a value to $v$, then the notation $v(E)$ for that value that $E$ assigns to $v$ is meaningful and well defined. 
Likewise, if $G$ is any register such that $E$ contains a write into $G$ then we denote with $G(E)$ the value of that write.
Again, since any operation execution contains at most one write action into any register, this notation is well defined.

In line 3, register \Th\ is read into variable $t$ and then the opposite  value is written onto register \Tp. 
So, the postman is always changing the color obtained from the homeowner process, while the homeowner always copy the
value obtained (see lines 3 and 4 in the \removeit\ code).  

In executing line 5, register \Rn\ is read into local variable $rn$, and in line 6 condition $rn < \itdn$ is checked.
If it holds then $1-t$ is written in \Fp, but otherwise the value $2$ is written. So $2$ is an indication that the mailbox
is empty.

There are two sorts of \checkit\ operations. A ``short'' \checkit\ $C$ is one that returns \TRUE\ immediately after line 1 is
executed. In this line, the homeowner process reads her own register \Fh\ and returns \TRUE\ if that register's value
is \TRUE. Note that line 1 is the only place in the algorithm where this register is read, and hence the register is
in fact dispensable and a local homeowner variable could replace it. The access restriction however prohibits 
persistent variables, and hence the need for this register which does nothing more than replacing a persistent local
variable.

A ``longer'' \checkit\ $C$ is one in which all lines 1 to 5 are executed. In lines 2, 3, 4
registers \Th, \Tp\ and \Fp\ are read, and the value that $C$ returns is a conjunction of two
statements that involve $\tp$, $\itth$,  and $\fp$. Note that \Tp\ and \Fp\ are registers of the postman process, 
but $\itth$ is the value of register \Th\ that the previous
\removeit\ operation determined or else is the initial value of that register (which is $0$) 
in case $C$ has no previous \removeit\ operation.

A \removeit\ operation is an execution $R$ of lines 1--6 of the \removeit\ code.
 First a letter is dequeued (and we have to prove that the queue
is nonempty when this instruction is executed) and then the current removal number $rn(R)$ is written on register \Rn.
 For any \removeit\ operation execution $R$, $rn(R)$ is the value of variable $rn$ after line 2 is executed in $R$.
 We have already noted that this notation is well defined since $rn$ is assigned a value in $R$ only at the execution of line 2.
  It follows
that $rn(R)$ is equal to $i$ where $R$ is the $i$-{th} \removeit\ operation execution. 
In lines 3 and 4 the homeowner copies the value read in register \Tp\ into register \Th. Register \Dn\ is then
read into $\itdn$ (line 5) and the boolean value  $``rn < \itdn  "$ is written in register \Fh.

The differentiation between a short and longer \checkit\ operations reflects a main idea of the algorithm, namely
that if the homeowner realizes in executing \removeit\ operation $R$ that $``rn < \itdn  "$ (namely that the queue is nonempty), then no subsequent 
postman operations can change this fact, and hence the first \checkit\ operation that comes after $R$ can rely on this information
and return \TRUE\ in a short execution.

There are two or three main ideas that shape our mailbox algorithm. The first one  (very roughly speaking) is that 
the inequality of registers $\Tp$ and \Th\ indicates a nonempty queue. Initially both registers are $0$, and in any \deliverit\
operation the postman reads \Th\ and writes in \Tp\
 a different value, thus indicating that the mailbox is nonempty. The homeowner
cancels this indication in any \removeit\ operation, but the equality of the values of registers
\Tp\ and \Th\ is not an assurance that the queue is empty. For example, after several letters were deposited, the homeowner
removes a single letter, leaving the two registers with equal value, and yet the queue is still nonempty. Of course,
registers \Rn\ and \Dn\ give an exact estimation of the number of letters in the mailbox (namely $\Dn-\Rn$), but since
the \checkit\ operation is not allowed to access these unbounded registers it has to rely on the bounded registers.
 The homeowner also
checks the boolean value \Fh\ and if it is \TRUE\ then the queue must be nonempty and the \checkit\ operation
is short in this case. (The queue is nonempty in this case because if the previous \removeit\ operation has established that 
$\Dn-\Rn>0$ then the mailbox {\em is} nonempty since no \removeit\ operations were executed between the
previous \removeit\ and the present \checkit.)  If, however, \Fh\ is \FALSE, the homeowner needs a more complex evidence
in order to deduce that the mailbox is nonempty: the inequality of colors $\tp\not = \itth$,  and the accordance 
$\fp = \tp $ (which also indicates that $fp\not = 2$).

\begin{figure}[H]
\label{label-of-the-figure}
\fbox{
\begin{minipage}[t]{16cm}
\linethickness{1.5pt}

\setlength{\unitlength}{1cm}
\begin{picture}(0,5)

%D_1
\put(0,3.5){\line(1,0){1.7}}
\put(0.7,3.7){$D_1$}

\put(0,3.4){\line(0,1){0.2}}
\put(1.7,3.4){\line(0,1){0.2}}

%C_1
\put(2,1.5){\line(1,0){1.7}}
\put(2.7,1.7){$C_1$}

\put(2,1.4){\line(0,1){0.2}}
\put(3.7,1.4){\line(0,1){0.2}}

%D_2(1-2)
\put(4,3.5){\line(1,0){1.7}}
\put(4.2,3.7){$D_2(1-2)$}

\put(4,3.4){\line(0,1){0.2}}
\dottedline{0.2}(4,3.5)(12,3.5)

%R_1
\put(6,1.5){\line(1,0){1.7}}
\put(6.7,1.7){$R_1$}

\put(6,1.4){\line(0,1){0.2}}
\put(7.7,1.4){\line(0,1){0.2}}

%C_2
\put(8,1.5){\line(1,0){1.7}}
\put(8.7,1.7){$C_2$}

\put(8,1.4){\line(0,1){0.2}}
\put(9.7,1.4){\line(0,1){0.2}}

%R_2
\put(10,1.5){\line(1,0){1.7}}
\put(10.7,1.7){$R_2$}

\put(10,1.4){\line(0,1){0.2}}
\put(11.7,1.4){\line(0,1){0.2}}

%D_2(3-6)
\put(12,3.5){\line(1,0){1.7}}
\put(12,3.7){$D_2(3-6)$}

\put(13.7,3.4){\line(0,1){0.2}}

%C_3
\put(14,1.5){\line(1,0){1.7}}
\put(14.7,1.7){$C_3$}

\put(14,1.4){\line(0,1){0.2}}
\put(15.7,1.4){\line(0,1){0.2}}

\end{picture}
\end{minipage}}
\caption{An example for an incorrect execution where a long check event only checks condition $tp\not = th$.}
\label{FIG}
\end{figure}
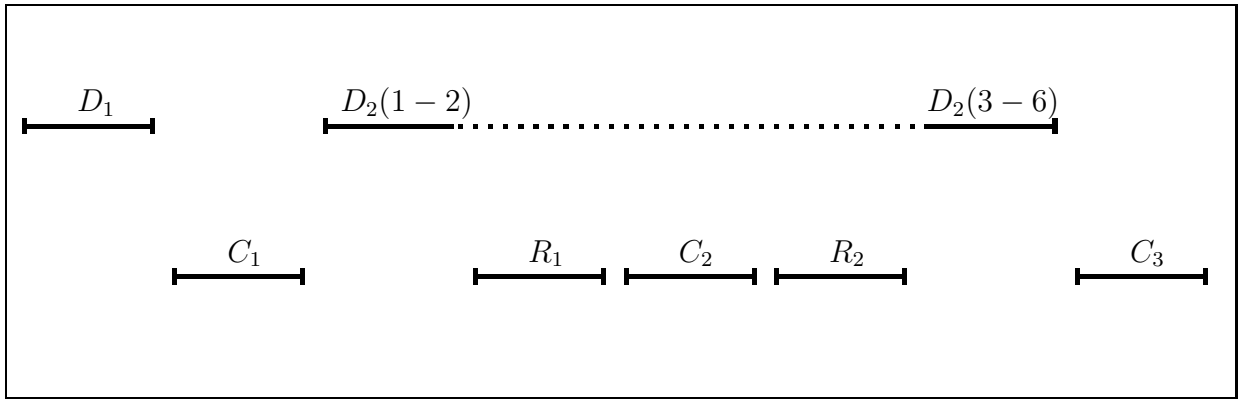

An example can be useful here to explain why this condition $\tp\not= \itth\ \wedge\ \fp=\tp$ cannot be replaced with 
the simpler condition  $\tp\not= \itth$. We see in figure \ref{FIG} the following course of events.

\begin{enumerate}
\item \postman\ execute a deliver operation $D_1$.
\item \homeowner\ execute a check operation $C_1$, since \postman has just delivered a letter, $C_1$ is longer and positive.
\item \postman\ starts to execute a second deliver operation D2 and execute the commands in lines 1 and 2. It sends the letter, writes 2 into register $\Dn$ and stops for awhile.
\item \homeowner\ execute a remove operation $R_1$. 
This is the first remove operation and \homeowner\ reads in this operation 2 from $D_num$ (the value that \postman\ wrote to $D_num$ in $D_2$). Hence, $R_1$ is positive.
\item \homeowner\ execute a check operation. Since $R_1$ is positive, $C_2$ is short and positive.
\item \homeowner\ execute a remove execution $R_2$. $rn(R_2)=2$ and $dn(R_2)=2$ (the value that \postman\ wrote to $\Dn$ in $D_2$). Thus, $R_2$ is negative.
\item $P_1$ completes the execution of $D_2$, and execute the commands in line 3-6. It reads a value
 $c$ from $T_H$ (this value has been written to $T_H$ during the execution of R2) and writes to register $T_P$, $1-c$.
\item \homeowner\ execute a check operation $C_3$. \homeowner\ reads the value $c$ from $T_H$ 
(written in the execution of $R_2$) and reads the value $1-c$ from register $T_P$ 
(written in the execution of $D_2$). Since only condition $tp\not =th$ is checked in $C_3$, $C_3$ is positive. 
Since there are only two deliver events and only two remove events in this execution, and since 
all of these executions precedes $C_3$, $C_3$ should be negative. Thus, this is an incorrect execution.
\end{enumerate}

\section{Correctness of the algorithm}
\label{Correctness}

In order to prove that our algorithm implements a mailbox (as specified in
Figure \ref{FL})  we need to define some functions and predicates that will serve us in this proof. An {\em action} is an execution of an atomic instruction of the algorithm such as a read or a write of a register or a queue action.
 Since we assume that the
registers are serial, and as the queue operations (to add or remove a letter) are also instantaneous, we have a total
ordering $<$ on these actions. We write $a<b$ to say that $a$ precedes $b$ in this ordering. (A relation $<$ is a total ordering
when it is a transitive and irreflexive relation such that for any two different members $a$ and $b$ 
 in its domain we have $a<b$ or $b<a$.) 

 An {\em operation execution} is an execution of the \deliverit, \checkit, or \removeit\ algorithm.
  Every operation execution is a high-level event, namely a set of lower-level actions (also called lower-level events, as in \cite{Lamport86}). The total ordering $<$ on the lower-level actions induces a partial ordering on the operation executions: for 
 operation executions $A$ and $B$ we define that $A<B$ if $a< b$ for every $a\in A$ and $b\in B$. It is also very convenient 
 to relate high-level events
and lower-level actions: $A<x$ for a high-level event $A$ and a lower-level event $x$ means that $a<x$ for
every $a$ in $A$. And similarly $x<A$ is defined when $x<a$ for every $a$ in $A$. The fact that we use the same symbol $<$ to
denote both the total ordering relation on the actions and the resulting partial ordering relation on the high-level events
should not be a source of confusion. 

The aim of the correctness proof is to define a total ordering $\prec$ on the operation executions that
extends the partial ordering $<$, and then to prove that the specifications of Figure \ref{FL} hold.

We assume two initial high-level events $I_p$ and $I_h$ by the \postman\ and \homeowner\ processes that determine 
the initial values of the registers (defined in Figure \ref{table}) and the initial values of the variables. 
$I_p$ contains the initial write actions on registers \Dn, \Tp, and \Fp, and $I_h$ contains the initial write
actions on registers \Rn, \Th\ and \Fh. These initial high-level events are concurrent. That is, it is neither the
case that $I_p < I_h$ nor that $I_h < I_p$.

If $a$ is any read/write action, then $[a]$ denotes that high level event to which $a$ belongs. (Every low level action belongs to some operation execution, except for the assumed initial write actions which belong to the initial events $I_h$ and $I_p$.)

We shall name the different actions that compose the three operations.
\begin{enumerate}
\item
Let $D$ be a \deliverit\ operation execution (which completed execution of lines 1--6 of the \deliverit\ code of Figure
\ref{fig:alg84}). We shall name the different actions of $D$. First, the addition of the letter to the queue $Q$ is denoted $\enq(D)$. $D$ contains three write actions denoted $w1(D)$ $w2(D)$ and $w3(D)$ (corresponding to lines 2, 4, and 6 respectively, namely the writes on registers $\Dn, \Tp$ and $\Fp$). $D$ contains two read actions $r1(D)$ and $r2(D)$ (which correspond to lines 3 and 5, namely to the reads of registers \Th\ and \Rn).

\item
There are two sorts of \checkit\ executions. A {\em short} operation $C$ is an execution of line 1 that returns the
value \TRUE. It contains a single read, denoted $r0(C)$, of register \Fh. A {\em longer} \checkit\ operation is one 
that contains executions of lines 1--5, and so it contains three additional read actions denoted $r1(C)$, $r2(C)$ and $r3(C)$. 
 $r1(C)$ is the read of register \Th, $r2(C)$ is the read of 
 register \Tp, and $r3(C)$ is a read of register \Fp. A \checkit\ operation contains no write actions.

  \item
A \removeit\ operation execution $R$ begins with a dequeue action on the mailbox queue $Q$ which is denoted
 $\deq(R)$. An important part of the correctness proof is to prove that whenever $\deq(R)$ is executed, $Q$ is nonempty. There are two read actions in $R$, $r1(R)$ and $r2(R)$ which correspond to lines 3 and 5. These are the reads of registers
 $\Tp$ and $\Dn$. Then we notate the three write actions: $w1(R)$ is the write on register \Rn,  $w2(R)$ 
is the write on register \Th, and $w3(R)$ is the write on \Fh.
\end{enumerate}

 If $X$ is a \deliverit\ (\removeit) operation execution, then $X$ contains a read action of the \Rn\ (respectively \Dn)
 register. Specifically, $r = r2(X)$ is the read of the \Rn\ (respectively \Dn) register, and then $\omega(r)$
 is the write action of that register which affected $r$. That is, $\omega(r)$ is the last write action on register
 \Rn\ (respectively \Dn) that precedes $r$ (see section \ref{PDqa}). Any action belongs to a unique higher level event, and if $Y$ is that
 higher level event that contains the write $\omega(r)$, then we define $Y=\alpha(X)$.
 
 A succinct definition of the function $\alpha$ can be given by the following equation. For any \deliverit\ or
 \removeit\ operation $X$ we define

\begin{equation}
\label{Eqone}
\alpha(X) =[ \omega(r2(X))].\end{equation}
  Recall  that $[a]$ denotes the higher level event that contains action $a$. In case $X=D$ is a \deliverit\ operation
  execution, $[ \omega(r2(D))]$ is that high-level event that contains $\omega(r)$, and so $\alpha(D)$ can
either be an operation execution that contains $\omega(r)$, or else the initial event $I_h$ of the \homeowner\
process in case $\omega(r)$ is the assumed initial write. 

In case $R$ is a \removeit\ operation execution, we have that 
$ \alpha(R) = [\omega(r2(R)]$. 
So if $r=r2(R)$ is the read of register \Dn\ in $R$, then $\omega(r)$ is the corresponding write action on that register.
We shall prove in Proposition \ref{P11}  that $\omega(r)$ is 
not the initial write in $I_p$, and so $D=\alpha(R)$ is a \deliverit\ operation execution and thus $\omega(r)=w1(D)$.  

The following lemma is an easy consequence of the fact that the registers (and specifically the \Dn\ register) are serial.
\begin{lemma}
\label{RR}
If $R_1<R_2$ are two \removeit\ operations, then $\alpha(R_1)\leq \alpha(R_2)$.
\end{lemma}

We say that a \checkit\ operation $C$ is ``positive'' in case it returns the value \TRUE. We say that it is ``negative''
when it returns \FALSE. Likewise, a \removeit\ operation $R$ is positive when it writes \TRUE\ on its \Fh\ register (in
executing line 6), and it is negative when it writes \FALSE. And, again, a \deliverit\ operation $D$ is positive if
condition $``\itrn < dn "$ holds at line 6 of $D$, and it is negative otherwise.

Now we define two functions, $\prerem$ and $\rho$, on the \checkit\ events.
\begin{definition}
\label{HN1}
Let $C$ be any \checkit\ operation execution.
 Define $\prerem(C)$ as the last \removeit\ operation execution $R$ such that $R<C$ if there is such a \removeit\ execution 
 that precedes $C$, and $\prerem(C)=I_h$ as the assumed initial \homeowner\ event otherwise.
 
 \end{definition}
  We note that a short \checkit\ operation is positive, and hence a \checkit\ operation $C$ is short if and only if $\prerem(C)$
  is positive. Since the assumed initial homeowner event is negative (as the initial value of
   \Fh\ is \FALSE), if $C$ is short then $\prerem(C)$ is not the initial event--
  it is necessarily a positive \removeit\ operation execution.

The following is a key definition in our correctness proof. It relates every \checkit\ operation $C$ to $\rho(C)$ which is the \deliverit\ operation (or initial $I_p$ event) that $C$ considers in order to calculate the value (\TRUE\ or \FALSE) to return.
\begin{definition}
\label{HN2}
For any \checkit\ operation execution $C$ we define $\rho(C)$ as follows. In case $C$ is a short \checkit\ operation let $R=\prerem(C)$ (which is a \removeit\ operation execution as we noted) and then
define $\rho(C)=[\omega(r2(R)) ]$. So $\rho(C)=\alpha(\prerem(C))$ when $C$ is short. In case $C$ is a longer operation, define  $\rho(C)= [\omega(r3(C))]$. ($r3(C)$ is the read of \Fp\ in $C$.)

\end{definition}

We note that $C<\rho(C)$ is impossible, by properties of the $\omega$ function (namely by the fact that $\omega(r)< r$ for any read action $r$). The following is therefore established.
\begin{prop}
\label{NP}
 If $C<D$ (where $C$ is a \checkit\ and $D$ a \deliverit\ operation) then $\rho(C)<D$.
\end{prop}

\begin{lemma}
\label{lemm1}
 Suppose that $C<R$ are a \checkit\ and \removeit\ operation executions. Then $\rho(C)\leq \alpha(R)$.
 \end{lemma}
 Proof. If $C$ is short then $\rho(C)=\alpha(\prerem(C))$, and since $R'=\prerem(C)<C<R$, $R'<R$ follows
 and so the proof is concluded in this case with Lemma \ref{RR}. Suppose next that $C$ is a longer
 \checkit\ operation and $\rho(C)=D$. Then $D= [\omega(r3(C))]$ by definition of $\rho$. This implies
 that $D<r3(C)$. (Because if $D=I_p$ is the initial event then $D<C$, and if $D$ is a \deliverit\
 operation then the fact that the write on \Fp\ is the last action in $D$ implies that $D< r3(C)$.)
 So $D<R$ and hence $D\leq \alpha(R)$.
 \qed

We remind the reader that if $E$ is any operation execution and $x$ a variable (or a register) whose value is assigned in $E$,
then $x(E)$ denotes this value.

For any \removeit\ operation $R$,   $rn(R)$ is the value of variable $rn$ that is determined in executing line 2 and
is written on register \Rn. We also set $rn(I_h)=0$ (and the initial value of variable $rn$ is $0$).

$rn(R)$ is called the ``removal number''; it is the number of \removeit\ operations $R'$ such
that $R'\leq R$. 
Clearly, if $R_1< R_2<\cdots$ is the sequence of \removeit\ operations in increasing order, then
$rn(R_i)=i$. 

The \checkit\ code does not contain a variable named $rn$, and so the number $rn(C)$ for a \checkit\ operation
execution $C$ is defined directly as the number of \removeit\ operations $R$ such
that $R < C$.
In other words, 
\begin{equation}
\label{RN}
rn(C) = \#\{R\mid R\ \text{is a \removeit\ operation and } R<C\}.
\end{equation}
Where $\#A$ denotes the cardinality of the set $A$.  

The ``delivery number'', $dn(D)$, of a \deliverit\ operation $D$ is equal to the number of \deliverit\
operations $D'$ such that $D'\leq D$. Thus if $D_1<D_2<\cdots$ is the enumeration of the \deliverit\ operations in increasing order, then $dn(D_i)=i$. We also define $dn(I_p)=0$.

It is convenient to define the ``color'' of operations. If $D$ is a \deliverit\ operation, then $color(D)=\Tp(D)=1-t(D)$.
That is, $color(D)$ is that value $c=0,1$ that is written into register \Tp\ when line 4 is executed in $D$. (If condition $rn<dn$ holds in line 6, then $color(D)$ is also
the value that is written in register \Fp.)

The color of any \removeit\ operation $R$ is defined by $color(R)=t(R)=\Th(R)$. That is, the color of $R$ is the value read from
register \Tp\ and written into \Th. The color of the initial event $I_p$ is $0$ which is the initial value of \Th.

If $C$ is a long \checkit\ operation, then we define $color(C)= tp(C)$. That is, the color of a long \checkit\ operation is the value read from register \Tp.

Note that if $C$ is a long
 \checkit\ operation and $D=\rho(C)$, if $C$ is positive then $D$ is positive and $color(D) = color(C)$. Indeed, $D=\rho(C)$
implies that the value read in register \Fp\ in $C$ (namely $fp(C)$) is the value written
by $D$. Hence this value is not 2 (because $C$ is positive and condition $tp=fp$ implies that
$fp=0,1$). Hence $rn < dn$ holds in $D$, and therefore $D$ is positive, and $color(D)=color(C)$ follows.

Note also that if $C$ is a long \checkit\ operation and $S=\prerem(C)$ (a \removeit\ operation or $I_p$), if $C$ is
positive then $color(C) \not = color(S)$. This follows from equality $tp\not = th$ which holds at line 5 if $C$ is positive.

We gather these observations into the following.
\begin{lemma}
\label{Q}
If $C$ is a long \checkit\ operation and  $S=\prerem(C)$, then $C$ is positive iff $\rho(C)$ is positive, 
$color(C) \not = color(S)$, and 
  $ color(\rho(C)) = color(C)$.
\end{lemma}

Our aim now is to prove some properties of the functions 
and predicates that we have defined above. These properties will be used to define a linear ordering (total ordering) $\prec$ on the operation executions and to prove that the properties of Figure \ref{FL} hold.

\begin{lemma}
\label{P}
Suppose that $C$ is a long \checkit\ operation and $D=\rho(C)$. If $S=\prerem(C)$ is a \removeit\ operation such that
$w2(D)<r1(S)$, then $C$ is negative.
\end{lemma}
Proof. Let $c=color(D)$ be, as we have defined above, the value written into register \Tp, and assume for a contradiction
that $C$ is positive. So $color(C) = color(D)$ by Lemma \ref{Q}.  Since $w2(D)<r1(S)$ are a write and read actions on register \Tp, $w2(D) \leq \omega(r1(S))$.
\begin{enumerate}
\item[Case 1.] 
$w2(D)=\omega(r1(S))$. This entails that $color(D) = color(S)$, and hence that $color(C) = color(S)$ which implies by Lemma \ref{Q} that $C$ is negative.

\item[Case 2.] $w2(D) < \omega(r1(S))$.  This implies that $D<D'$ where $D'=[\omega(r1(S))]$, and $\omega(r1(S))=w2(D')$.
So  $w2(D') < r1(S)$ (as $\omega(r)<r$ for every read action $r$). 
Since it is not the case that $w3(D')< r3(C)$ (as $\rho(C)=D$), we get that  $r3(C)<w3(D')$. Hence $w2(D')<r1(S)< r2(C)< w3(D')$. This implies that $r1(S)$ and $r2(C)$ (which are both reads of register \Tp) get the same value of the write $w2(D')$. Hence $color(S)=color(C)$ which implies, again by Lemma \ref{Q}, that $C$ is negative.
\end{enumerate} 
\qed

\begin{prop}
\label{P11}
If $C$ is a positive \checkit\ operation and $\rho(C)=D$, then $D$ is a \deliverit\ operation execution and
\begin{equation}
\label{EG}
 rn(C) < dn(D).
\end{equation}
 
\end{prop}
Proof. Assume first that $C$ is a short \checkit\ operation and let $R=\prerem(C)$ be the previous \removeit\
operation, which necessarily has set its register \Fh\ to be true at line 6. So $rn(R) = rn(C)$, and inequality 
\begin{equation}
\label
{IE} rn(R) < dn(R)
\end{equation} 
holds. Let $r=r2(R)$ be the
read of register \Dn\ which obtained the value $dn(R)$. By definition of $\rho(C)$ when $C$ is short, $D =\rho(C)=[\omega(r)]$,
and $dn(D)=dn(R)$ follows. Since $dn(R)>0$ follows from (\ref{IE}) and as $dn(I_p)=0$, $D\not = I_p$ is concluded and necessarily
$D$ is a \deliverit\ operation execution and  (\ref{EG}) follows.

Now suppose that $C$ is a longer \checkit\ operation, and let $r2=r2(C)$ and $r3=r3(C)$ be its reads of registers \Tp\ and
\Fp\ (respectively). By definition of $D = \rho(C)$, $D=[\omega(r3)]$. Then $D$ is either  a \deliverit\ operation
execution (in which case $\omega(r3)=w3(D)$ is the write in register \Fp)
 or else is the initial event $I_p$ in case $\omega(r3)\in I_p$. We claim that $D$ is not the initial
 event $I_p$. Indeed, the initial value of $F_p$ is $2$, but as $C$ is positive condition $fp=tp$ holds in $C$,
 which excludes the possibility that $fp(C)=2$ (as $tp(C)\in\{0,1\}$). Hence $fp(C)=fp(D)$ is
 not $2$ and so $ \itrn < dn $ is evaluated to \TRUE\ when line 6 is executed in $D$. So \[rn(D)< dn(D)\] holds.
   
 Define $R=\alpha(D)$; that is $R=[\omega(r2(D))]$. Then $rn(D)=rn(R)$. We shall prove that $R=\prerem(C)$. This will show that $rn(C)=rn(R)$, and hence that $rn(C)< dn(D)$ as required. It thus remain to prove that $R=\prerem(C)$. 
 
 Suppose on the contrary
 that $R\not = \prerem(C)$, and then $R<\prerem(C)$ follows (from the fact that $w1(R)<r2(D)<w3(D)<r3(C)$ which implies that
 $R<C$). Say $S=\prerem(C)$. Since $\omega(r2(D))$ is in $R$, $r2(D)< w1(S)$. But $w2(D)<r2(D)$. Hence $w2(D)< w1(S) <r1(S)$ and this implies by Lemma \ref{P} that $C$ is not positive, which yields a contradiction.\qed
 
 \ignore{
 \begin{prop}
\label{P1}
If $R$ is a \removeit\ operation and $D=\alpha(R)$, then $dn(D)\geq rn(R)$.
\end{prop}
Proof.  Recall that the homeowner never approaches her mailbox unless a \checkit\ operation returns with a positive answer.
 Let $C$ be that positive \checkit\ operation that precedes $R$. Say $D'=\rho(C)$. Then $D'\leq D$ (by Lemma \ref{lemm1}). By the previous proposition $dn(D')>rn(C)$. But $rn(R)=rn(C)+1$. Hence $dn(D')\geq rn(R)$, and hence $dn(D)\geq rn(R)$.\qed

\begin{prop}
\label{P2}
 For every \removeit\ operation execution $R$, if $D=\alpha(R)$, then $R$ is positive if and only if $rn(R) < dn(D)$.
\end{prop}
Proof. By definition of $D=\alpha(R)$, the read $r2(R)$ (of register \Dn\ in $R$) obtains the value that $w1(D)$ writes.
Hence $dn(R)=dn(D)$. Now, $R$ is positive if and only if $rn(R) < dn(R)$, and hence the proposition follows. \qed
}%%%%%%%%%%%%%%
\begin{prop}
\label{P3}
If $D<C$ are a \deliverit\ and \checkit\ operations such that $rn(C)<dn(D)$, then $C$ is positive.
\end{prop}
Proof. A short \checkit\ operation is always positive (it returns \TRUE), and hence we may assume that $C$ is a longer \checkit. 
Say $R=\prerem(C)$ and then
\begin{equation}
\label{ERN}
rn(R) = rn(C).
\end{equation}
Suppose first that $R=I_h$ is the initial event. 
 In reading \Th, $C$ obtains
the initial value $0$. We shall prove that $fp(C)=tp(C)=1$, and hence that $C$ returns \TRUE\ at line 5, as required.

 Define $E=\rho(C)=[\omega(r3(C))]$. Then $w3(E)<r3(C)$ (the write on \Fp\ in $E$ precedes the read of this register
 in $C$), and $D<C$ implies that $D\leq E$.  Now
  $\alpha(E)=I_h$ follows from the assumption that $\prerem(C)=I_h$. $rn(E)$ is $0$ (the initial value of \Rn), but $dn(E)>0$. So $``rn<dn"$ holds in $E$ when line 6 is executed in $E$, and hence  the value of $w3(E)$ is
 $1-t(E)$. But $t(E)=0$ because the initial value of \Th\ is $0$, and hence the value of $w3(E)$ is $1$. So $fp(C)=1$. 
 The proof that $tp(C)=1$ is very similarly obtained by taking $[\omega(r2(C))]$. 

So now we assume that $R$ is a \removeit\ execution. In case $w1(D) < r2(R)$, 
$w1(D)\leq \omega(r2(R))$ follows, and hence the read of \Dn\ in $R$ obtains the write in $D$ or a later write.
Hence $dn(D)\leq dn(R)$.
 The fact that $rn(R)=rn(C)$ and our assumption
that $rn(C)<dn(D)$ imply that $rn(R)<dn(R)$. So $R$ is positive and $C$ is a short positive \checkit\ operation.

 So we may assume that $r2(R)<w1(D)$.
 It follows from this assumption that $w2(R)<r1(D)<C$.
   Say $c = color(R)$ (that is, by definition, the value of $w2(R)$, which
is the write in \Th).

\noindent
{\bf Claim.}
If $E$ is any \deliverit\ operation such that $w2(R)<r1(E)<r3(C)$ (the write on \Th\ in $R$ precedes the read of \Th\ in $E$ which itself precedes the end of $C$) then $\omega(r1(E))= w2(R)$ and $color(E) = 1- c$. 
 
 \noindent
{\bf Proof of claim.} Since  $r1(E)$ is before the end
of $C$ there is no write action on register $\Th$ between $w2(R)$ and $r1(E)$. Hence $w2(R)=\omega(r1(E))$. So
$color(E) = 1- c$.

 In particular, if $E_0=\rho(C)$, then $D\leq E_0$ and the conditions of the claim hold. (Recall that
 $r3(C)$ is the read of register \Fp\ in $C$, and $\rho(C)=[\omega(r3(C))]$.
 Since $D<C$, $D\leq E_0$. And as the write on \Fp\ is the last action in $E_0$, $E_0< r3(C)$.)
  Thus $color(E_0)=1-c$.
 Moreover, $R=\alpha(E_0)$. To prove this fact note that $w1(R)<w2(R)<r1(E_0)<r2(E_0)$ and $r2(E_0)$ is before the end of $C$; this implies that $w1(R)=\omega(r2(E_0))$ and hence that $rn(E_0)=rn(R)$. But  (\ref{ERN}) and the lemma's assumption
 give $rn(R)=rn(C)<dn(D)$, and since $D\leq E_0$ yields $dn(D)\leq dn(E_0)$, condition $rn(E_0)<dn(E_0)$ holds.
Hence the value of \Fp\ that is written by $E_0$ is the color of $E_0$ which is $1-c$. Since $E_0=\rho(C)$, this implies that $fp(C) = color(E_0)=1-c$, and thus \[fp(C) = 1-c.\]

Condition $fp = tp$ holds in $C$ by the following argument. 
 $tp(C)$ is the value of the read  of \Tp, namely the value of 
$r2(C)$. Say $E=[\omega(r2(C))]$, that is $w2(E)=\omega(r2(C))$. Since $D<C$, $D\leq E$. Also, $r1(E)$ is before the end of $C$. As we noted in the above claim, this implies that $color(E)=1-c$, and hence \[tp(C)= 1-c.\] In view of the formula displayed above, this yields
that $fp = tp$ holds in $C$.

Since $color(R)=c$, $R$ writes $c$ on \Th. But $R=\prerem(C)$, and so $th(C)=c$ follows.
Hence condition $tp\not= th$ holds in $C$ because $tp(C)=1-c$ but $th(C)=c$. So $C$ is indeed positive. \qed

We are now ready to define the linear ordering $\prec$ on the \deliverit, \checkit\ and \removeit\ operations. 
We shall define first a relation $<^*$ that extends $<$ on the operation executions, and then prove that $<^\ast$ has no cycles,
and that any linear ordering $\prec$ that extends $<^*$ satisfies the linear mailbox specifications of Figure \ref{FL}. This will complete the proof.  We define the relation
$<^*$ as a union of $<$ with the relation $\lhd$ that relates some \checkit\ operations $C$ and \deliverit\ operations $D$ as follows.

\[
\begin{array}{rcl} \lhd & = & \{ \langle C,D\rangle\mid C\ \text{is negative and } rn(C)< dn(D)\}  \\

& & \hspace{6mm} \cup \{ \langle D,C\rangle\mid C\ \text{is positive
 and } dn(D) = rn(C)+1\}.
 \end{array}\]

 Before we proceed we want to explain the intuition behind this definition of $\lhd$.
If $C$ is a negative \checkit\ operation and $rn(C) < k$, if $D$ is the $k$th \deliverit\ operation or a later \deliverit,
then we surely want to have $D$ after $C$ in the linear ordering $\prec$ that we look for. (Otherwise, if $D$ is before $C$,
 then $C$ is required to be positive.) If, on the other hand, $C$ 
is positive then among the operations that are before $C$ in the $\prec$ ordering we must have more \deliverit\ than 
\removeit\ operations and hence the $k+1$ \deliverit\ operation must be before $C$.  

\begin{lemma}
\label{L1}
If $X\lhd Y$ then it is not the case that $Y<X$.
\end{lemma}
Proof. We have to check two cases as in the definition of $X \lhd Y$.
\begin{enumerate}
\item Suppose first that $\langle X,Y\rangle = \langle C,D\rangle$ where $C$ is a negative \checkit\ operation and $D$ is a \deliverit\ 
operation such that  $rn(C)< dn(D)$.
We have to prove that it is not the case that $D<C$. But if $D<C$ then Proposition \ref{P3}
implies that $C$ is positive.

\item Suppose next that $\langle X,Y\rangle = \langle D,C\rangle$ where $C$ is a positive \checkit\ 
operation and $D$ is a \deliverit\ operation such that $dn(D)=rn(C)+1$. We have to prove that
it is not the case that $C<D$. But if $C<D$, then $D'=\rho(C)<D$ (by Proposition \ref{NP}) and 
hence $dn(D')<dn(D)$. So $dn(D')\leq rn(C)$, in contradiction
to Proposition \ref{P11}.

\end{enumerate}\qed 

\begin{lemma}
\label{L3}
If $C$ and $C'$ are \checkit\ operations and $D$ is a \deliverit\ operation such that $C\lhd D \lhd C'$, then $C<C'$.
\end{lemma}
Proof. Since $C\lhd D$, the definition of $\lhd$ implies that $C$ is negative, and \[rn(C)<dn(D).\] Now from $D\lhd C'$ we get that $C'$ is
positive and $dn(D)= rn(C')+1$. So, firstly, we infer that $C\not = C'$ (one is negative and the other positive).
If it is not the case that $C<C'$, then $C'< C$ holds. In this case, since $C'$ is positive, there is a \removeit\
operation between $C'$ and $C$, and hence $rn(C')<rn(C)$. So, $rn(C')<rn(C)< dn(D)$ which is in contradiction to  $dn(D)=rn(C')+1$.\qed

\begin{lemma}
\label{L4}
If $D$ and $D'$ are \deliverit\ operations and $C$ a \checkit\ operation, then $D\lhd C \lhd D'$ is impossible.

\end{lemma}
Proof. $D\lhd C$ implies that $C$ is positive but $C\lhd D'$ implies that it is negative.\qed

A cycle of length $k\geq 1$ in a relation $T$ is a sequence $X_1,\ldots,X_{k+1}$ so that $X_i T X_{i+1}$ for $1\leq i\leq k$, and
$X_{k+1} = X_1$. We say that $X_{i+1}$ is the successor of $X_i$ in this cycle.
\begin{lemma}
\label{L5}
Relation $<^*\; = (< \; \cup\; \lhd)$  has no cycles, and hence can be extended to a linear ordering of the operation
executions.
\end{lemma}
Proof. By the definition of the union of two relations, $X<^\ast Y$ if $X<Y$ or $X\lhd Y$.
 Suppose on the contrary that there is a cycle $X_1<^* X_2<^*\cdots <^* X_n$ of length $n\geq 1$ in the $<^\ast$ relation.
Take such a cycle of minimal length. Since $<$ is transitive, there are no two successive occurrences of the $<$ relation
in this minimal cycle. But it is also impossible to have two successive occurrences of the $\lhd$ relation (by lemmas \ref{L3}
and \ref{L4}). The cycle is not of length one, since both $<$ and $\lhd$ are irreflexive. The cycle is not of length two (use 
Lemma \ref{L1} to see that it is not of the form $X\lhd Y<X$ or $X<Y\lhd X$). 

We may assume that the cycle begins with the $<$ relation,
and so it begins $X_1< X_2\lhd X_3 < X_4 \cdots$. But $X_2\lhd X_3$ implies (by Lemma \ref{L1})  that it is not the case
that $X_3< X_2$. So  $\Begin(X_2)<\End(X_3)$, where $\Begin(X)$ and $\End(X)$ are the first and last actions in $X$. Hence
$X_1<X_4$ follows in contradiction to the minimality of the cycle. \qed

As $<^\ast$ has no cycles it can be extended to a linear ordering.
\begin{theorem}Let 
$\prec$ be any linear ordering (total ordering) that extends $<^*$. 
Then the specifications of Figure \ref{FL} hold.
\end{theorem}
Proof.  
For any \checkit\ operation $C$ we define $\Val(C)= \TRUE$ if $C$ is a positive, and $\Val(C)=\FALSE$ when $C$ is negative.
We now check the four items of Figure \ref{FL}.
\begin{enumerate}
\item
$\prec$ is chosen to be a linear ordering that extends $<^\ast$, and hence it also extends the $<$ ordering on the operation executions. We want to show that for every operation execution $X$ the set $\{Y\mid Y\prec X\}$ is finite. This is a consequence of the finiteness property of the $<$ relation which says that for every event $X$ there is only a finite number of events $Y$ such that $X<Y$ does not hold. Hence for all but a finite number of events $X\prec Y$ holds.

\item If $R$ is any \removeit\ operation, then $R$ is preceded by a positive \checkit\ operation $C$. This is a requirement on how the operations are invoked, and since the \homeowner\ process is a serial process the two ordering $<$ and its extension $\prec$ agree on the operations of that process, and so there is no \checkit\ or \removeit\ operation execution $X$ with $C\prec X\prec R$.

\item 
Recall that for every
\checkit\ operation execution $X$, $\delnum(X)$ and $\remnum(X)$ are the number of \deliverit\
operations $D$ such that $D\prec X$, and (respectively) the number of \removeit\ operations $R$ such
that $R\prec X$. We have defined (in (\ref{RN})) the number $rn(C)$ as the number of \removeit\ operations
$R$ such that $R<C$. Since the homeowner process is serial, relations $<$ and $\prec$ coincide on the homeowner
events, and hence 
\begin{equation}
\label{D1}
rn(C) = \remnum(C).
\end{equation}
And similarly, for any \deliverit\ $D$
\begin{equation} 
\label{D2}
dn(D)= \delnum(D).
\end{equation}
We have to show that 

 \begin{equation}\label{El} \Val(C)=``\remnum(C)<\delnum(C)".
 \end{equation}
 (Where $``\varphi"$ is the truth value of $\varphi$.)
Consider first the case that $C$ is negative, and assume that in contradiction to (\ref {El})  $\remnum(C)<\delnum(C)$.
Say $\remnum(C) = k$. So 
\begin{equation}
\label{ES} k < \delnum(C).
\end{equation} If $D_1<D_2\cdots$ is an enumeration in increasing $<$ order of the \deliverit\ operations, then $D_{k+1}\prec C$ 
(for otherwise, as $\prec$ is a linear ordering, $C\prec D_{k+1}$ and hence \[\{ D\mid D\ \text{is a \deliverit\ operation and } D \prec C\}\subseteq \{D_1,\ldots, D_k\}\] which implies that $\delnum(C)\leq k$ in contradiction to (\ref{ES})).  Yet, as $C$ is negative, $rn(C)=k$ and $dn(D_{k+1})=k+1$, the definition of $\lhd$ dictates that $C\lhd D_{k+1}$, which is in contradiction to $D_{k+1}\prec C$.

Consider now the case
that $C$ is positive. Say $D = \rho(C)$. By Proposition \ref{P11}, $rn(C) < dn(D)$.
Hence we do have a \deliverit\ operation $D$ with $dn(D)=rn(C)+1$. Then $D\lhd C$ and hence $D\prec C$. This shows
that $\delnum(D)\leq \delnum(C)$. But $rn(C)<dn(D)$ and equations (\ref{D1}) and (\ref{D2}) show that $\remnum(C)<\delnum(D)$
and hence that (\ref{El}) holds.

\item The fourth property of Figure \ref{FL}  is that $R_i$ obtains the letter of $D_i$. Let $C$ be that positive \checkit\
operation that precedes $R_i$. Then $rn(C)=i-1$. Define $D=\rho(C)$. By Proposition \ref{P11}, $rn(C) < dn(D)$. Hence
$dn(D)\geq i$. So \[D_i\leq D.\] This implies that $\enq(D_i)< \deq(R_i)$ (see below) and since this relation holds
for every $i$ and as we assume that the queue $Q$ that the algorithm employs is a fifo queue, it follows that the value
dequeued by $R_i$ is the value enqueued by $D_i$. Why $\enq(D_i)< \deq(R_i)$? If this is not the case and
$\deq(R_i) < \enq(D_i)$, then the fact that the enqueue action is the first in any \deliverit\ operation yields (together 
with $C<R_i$) that $C< D_i\leq D$. But $C<D$ is in contradiction to $D=\rho(C)$.  

\end{enumerate}

\section{A note on the proof}
Our correctness proof of the linearizability of the mailbox algorithm that was given in the previous section is clearly divided into two parts. The first part consists in defining relations and functions such as $\alpha$ and $\rho$, and in proving properties
of the operation executions that are expressed by means of these relations and functions. This part of the proof is
extended from Lemma \ref{RR} to  Proposition \ref{P3} and it relies on the text of the algorithm. The proof in the second part defines the linearization ordering $\prec$ and shows that
it possesses the required properties (those that are displayed in Figure \ref{FL}). In this part, the algorithm is not
mentioned and only properties established in the first part are used in an abstract way. Although the proof of both parts was
quite detailed and (we hope) convincing, we cannot claim that it is a formal proof because something very definite is lacking which
we want to explicate. The correctness condition (linearizability) is about executions of the algorithm, but we never defined
what these executions are; we never defined mathematical objects that represent executions and so
we did not explicate in a precise way how to formulate and formally  prove theorems about executions.

The standard way to define executions of a distributed algorithm is the following which is based on the notions of states, steps and runs.
A state is, informally speaking, a description of the system
as if frozen at a certain moment. Formally, a state is a function that assigns values to the state variables.  Variables of our system are, for example,  $PC_p$ (the postman program counter)
which can take any of the values in $\{1,\ldots,6\}$, $PC_h$ (which is the homeowner program counter), $T_H$ (which is
the register with values in $\{0,1\}$) etc. If $S$ is a state and $x$ is any of the state variables, then $S(x)$ denotes
the value of $x$ in state $S$. An initial state is a state $S$ such that $S(PC_p)=1$, $S(\Dn)=0$, and so on as in Figure \ref{table}.

A {\em step} is a pair of states $(S,T)$ that represents an execution of an (atomic) instruction by one of the processes.
So, for example, a ``read of register \Th'' by the postman process is a step $(S,T)$ such that $S(PC_p)=3$, $T(PC_p)=4$,
$T(t_p)=S(\Th)$ and for any variable $x$ different from $PC_p$ and $t_p$ $T(x)=S(x)$.

A {\em run} is defined to be a sequence of states $S_0,\ldots$ such that $S_0$ is an initial state and for every $i$
$(S_i,S_{i+1})$ is a step by one of the processes. Runs represent executions of the algorithm.

These runs cannot support the lemmas and propositions of the first part of our linearization proof and certainly they
do not suffice for its second part, simply because the high level events, namely the operation executions, are not an integral
 part
of these runs. Proposition \ref{P11} for example, requires the notion of \checkit\ and \deliverit\ operations, as well as
the functions $\alpha$, $dn$ and $rn$. Now, incorporating these higher level events and functions is nothing very deep. We can simply take a run with its actions (formed by the steps)  and define sets of actions that form the operation executions. This yields a structure that contains both
actions and higher level events, and the functions $\alpha$, $\rho$, etc. can be defined in this resulting structure as we did in the previous section. A detailed description of this process by which the extended run structure is obtained may be quite long, but
it is quite straightforward. In fact, there are possibly more then one reasonable way to achieve this construction and a particular one can be found in \cite{Bakery} and \cite{book}.

If we denote with $H$ some run of the system, that is some sequence of states $H=(S_0,\ldots)$ so that every pair $(S_i,S_{i+1})$
is a step, and if we let $\overline{H}$ be the resulting extended structure that contains both the actions, the higher level operation
executions and the required functions, then all the lemmas and propositions of the first part of our proof refer to the
structure $\overline{H}$ (or more correctly to the set of all structures $\overline{H}$ obtained from runs $H$ of the system).

Now for the second part of the proof we no longer need the actions and  references to the algorithm instructions.
The structures that interest us are those obtained by forgetting all references to lower level actions and keeping only
the higher level operation executions and the required functions and relations that are defined over them.  Let $\overline{H}$
be the  extended structure that results from a run $H$. Then we can form a structure $M$ by keeping only the operation
executions (as members of the universe of $M$), the precedence relation $<$ over these members and all functions
and predicates that are defined over them. The resulting structure $M$ is the one on which the second part of our linearization
proof is about. $M$ is a structure in the standard sense that is given in mathematical logic books. It is an interpretation
of some definite relational language. Any structure $M$ obtained in this way satisfies the properties that were established
in Propositions \ref{P11} to \ref{P3} and some additional obvious properties, and the second part of the correctness proof
establishes that any structure that satisfies these properties possesses a linearization as required by the 
linear mailbox specification of Figure \ref{FL}. We refer to structures such as $M$ as Tarskian system executions\footnote{This
term was chosen in order to indicate that we incorporate here the notion of system execution defined by Lamport \cite{Lamport86}
with the work and ideas of Alfred Tarski.}.

A careful reader would surely not be happy with our ``additional obvious properties'', and she would rightly 
request a more detailed definition.
What is needed (for a careful correctness proof) is a definition of a first-order language $L$ and a list of properties $PL$
that include not only those enunciated by the propositions but also all those additional properties that are required for the proof. Then the fact that the structures $M$ are detached from the algorithm help us to check that indeed only the assumptions
made in the list $PL$ (and all of these properties) are used in the second part of the proof. In our experience, this separation
of the correctness proof into two parts with the corresponding separation of the modeling structures helps to improve the algorithms whose correctness we try to prove. What often happens is that when the second part of the proof is established and
it is evident that only the properties listed in $PL$ are needed, then the algorithm itself can be changed and improved by
the designer who knows that if only these properties of $PL$ still hold then the algorithm is correct. 

To give an idea of what we have in mind for the list $PL$ we spell out in details such a list, but we first describe the language to which the statements of this list belong.
The $L$ language is a multi-sorted language that contains the following elements.
\begin{enumerate}
\item There are two sorts: \Event\ and \Number. (The role of sort \Event\ is to represent the
operation executions, and the role of \Number\ is to represent the set of natural numbers.)

\item The following unary predicates are defined over \Event. \deliverit, \checkit, \removeit,
\positive\ and \negative.

\item A binary relation $<$ is defined on the \Event\ sort. (This is called the precedence relation.) The same symbol
$<$ is also used for the ``smaller than'' relation on the \Number\ sort. The successor function $x+1$ is also assumed here.

\item The functions $rn$ and $dn$ are defined over the \Event\ sort and they take \Number\ values.

\item The function $\rho$ is defined on the \Event\ sort and with values in this sort. (In fact, we are only
interested in $\rho(C)$ when $C$ is a positive \checkit\ event, and in this case $\rho(C)$ is a \deliverit\ event.)
\end{enumerate}
  
The $PL$ properties are defined to be the following ``axioms''. (For simplicity we did not introduce queue events and
did not relate the \deliverit\ and \removeit\ events to the queue events.) 

\begin{enumerate}
\item Relation $<$ is irreflexive and transitive on the \Event\ sort, and it satisfies the following property\footnote{This
is the Russell--Wiener property which characterizes interval orderings.}.
\begin{enumerate}
\item
For every \Event\ members $X_1,X_2,X_3,X_4$:
\[ \text{if } X_1<X_2,\ X_3<X_4\ \text{and } X_2,X_3 \text{ are incomparable in } <, \text{ then } X_1<X_4.\]

\item For every event $A$ there is a finite set of events $F$ such that if $Y$ is any event not in $F$ then $X<Y$.
\end{enumerate}

\item 
The \deliverit, \checkit, and \removeit\ predicates are disjoint. We write $\homeowner(x)$ for $\checkit(x)\vee \removeit(x)$.

\item The \deliverit\ events are linearly ordered. That is, if $\deliverit(e_1)$ and $\deliverit(e_2)$, if $e_1\not = e_2$, then
$e_1< e_2$ or $e_2< e_1$. 

The function $dn$ is an enumeration of the \deliverit\ events in their ordering. That is,
for every \deliverit\ event $d$, $dn(d)$ is the number of \deliverit\ events $d'$ such that $d'\leq d$.
(So $dn$ is one-to-one, into \Number\ and with values $>0$, so that for every
\deliverit\ events $d_1$ and $d_2$ $dn(d_1)<dn(d_2)$ iff $d_1<d_2$, and if $dn(d)=k$ then for every $1\leq j <k$ there
exists some \deliverit\ $d'$ with $dn(d')=j$.)

\item The \homeowner\ set of events is linearly ordered,
and if $\homeowner(x)$ then $rn(x)$ is the number of \removeit\ events $r$ such that $r\leq x$.

\item We assume an initial event $I$ and $I<e$ for any other event $e$. 
 
\item Any \checkit\ event is either positive or else negative.
If $C$ is a positive \checkit\ event then there exists some \removeit\ event $R$ such that
$C<R$ and there is no \homeowner\ event $X$ with $C<X<R$. 

If $R$ is a \removeit\ event then there is some positive \checkit\ $C$ such that $C<R$ and there is no \homeowner\ event $X$ with $C<X<R$. 

\item  If $C$ is a positive \checkit\ event and $\rho(C)=D$, then $D$ is a \deliverit\ operation and $rn(C)<dn(D)$.

\item If $D<C$ are a \deliverit\ and (respectively) a \checkit\ events such that $rn(C)<dn(D)$ then $C$ is positive.

\item If $C<D$ are a positive \checkit\ and (respectively) a \deliverit\ events, then $\rho(C)<D$. 
\end{enumerate}
The last three items, 7,8 and 9, are the main properties and they were established in propositions \ref{P11}, \ref{P3} and
\ref{NP}.
The reader can return now to section \ref{Correctness} and re-read the second part of the proof, but now as if it were
an abstract proof about arbitrary structures that posses the nine properties listed above. The reader can check that
indeed only these properties are used in the proof and each one serves at some point. (The argument that involves the 
\Begin\ and \End\ functions can be adapted to one the employs the Russell--Wiener property.)

The role of the function $\rho$ is intuitively evident. If $C$ is a positive \checkit\ operation then it must be the
case that $C$ relies on some \deliverit\ operation execution $D$ that ensured $C$ that it may return \TRUE. The function
$D=\rho(C)$ gives this assurance, based on the inequality $dn(D)>rn(C)$. And of course, we cannot expect that $C$ relies
on some future event: hence $C<\rho(C)$ is ruled out.
It is not difficult to check that $\rho$ is not only intuitively appropriate, but it is in fact 
necessary in the sense that if we do have a mailbox
algorithm for which a linear ordering $\prec$ exists that satisfies the condition of Figure \ref{FL} then a function
$\rho$ can be defined that satisfies items 7 and 9.

\section{Conclusion}

In \cite{AGL2},  Aguilera, Gafni, and Lamport define the Mailbox problem, and present a solution in which the
\checkit\ operation reads two registers (the ``flag'' registers) that can carry 14 values each. Moreover,
they prove that there is no solution to the Mailbox problem with two binary flags. We have presented
here a much simpler solution to the Mailbox problem with two flags that can carry 6 and 4 values each.
The gap between the impossibility of solving the Mailbox problem with binary flags and our solution with flags that
have 10 values in total is meaningful and it poses interesting theoretical questions: to
improve on the lower bound of \cite{AGL2}, and to find a better solution to the Mailbox problem than the one
presented here. 

Another problem from \cite{AGL2} is whether the space efficiency of the mailbox algorithm presented in that paper can be improved.
The algorithm of \cite{AGL2}  uses
$\Theta(n \log n)$ bits of shared memory, where n is the number of executions of
deliver and remove. The authors of \cite{AGL2} conjecture that there is a solution using logarithmic
space, and indeed our algorithm uses two registers \Dn\ and \Rn\ of width exactly $\log n$ for $n$ executions.

An
interesting problem (connected with the Mailbox problem) is posed in \cite{AGL2}: the bounded, wait-free Signaling problem for which \cite{AGL2} gives only a non-blocking solution and leaves the wait-free problem
open.
The ideas developed in this paper have contributed to a solution of the wait-free Signaling
problem which was obtained by the second author. 

There are other problems around the Mailbox problem that seem to be quite interesting.
Are there solutions to the mailbox problem in which all registers (not only the flag registers) are
bounded?
What solutions to the mailbox problem can be obtained in which the flags are simple registers but the
other registers and queues can be more complex shared memory devices (for example queues that have consensus number 2).

The last section of our paper discusses the structure of the correctness proof and outlines 
a more abstract, two-stage proof in which the first stage investigates the algorithm and the resulting
behavior of the higher level operation executions, and the second stage deals with abstract properties
that are detached from the algorithm's text. In our experience, this division of the correctness proof
into two distinct parts has some marked benefit that justifies further investigation.
Not only that the correction proof seems clearer in our eyes when its two parts are thus formally
delineated, but the method helps to fashion better algorithms. In developing the algorithm there is
a stage when the second part of the proof (its higher level, abstract part) is established but the
algorithm itself is not yet completely determined; there are some features in the algorithm that can
still be changed, some actions that can be omitted, and some data structures that can be reduced. 
When the designer of the algorithm has a clear and accessible aim in mind, namely when the higher level
properties that the algorithm has to ensure are written down, then this process of improving the design of
the algorithm follows a sure path. For example, in the process of designing the mailbox algorithm, once
we understood that it suffices for the algorithm to satisfy the nine properties listed above in order to
solve the mailbox problem we could play with changes and improvements knowing that as long as propositions
remain correct we are on the right path.

\end{document}